# Reclassification of thermal equilibrium phase transitions in thermodynamic limit systems

Lai Wei[1], Li-Li Zhang[1*], and Yi-Neng Huang[1,2*]

[1] Xinjiang Laboratory of Phase Transitions and Microstructures in Condensed Matters, College of Physical Science and Technology, Yili Normal University, China.

[2] National Laboratory of Solid State Microstructures, School of Physics, Nanjing University, China.

*Corresponding author. Email: suyi2046@sina.com; ynhuang@nju.edu.cn

## Abstract

For relaxor-ferroelectrics and relaxor-ferromagnets, Ehrenfest classification of phase transitions based on the discontinuity of entropy or specific heat etc. with temperature ($T$) gives no transition that contradicts the measured order parameter, i.e. spontaneous polarization and magnetization, at low temperatures, while Landau classification based on the minimum derivative of the order parameter to $T$ raises the question about the relationships between the phase transition and the specific heat peak above and near the transition temperature. Here, based on the free energy ($A|_{F\to 0}$) of the thermodynamic limit systems with $T$ when the external field ($F$) tends to 0, thermal equilibrium phase transitions are reclassified into three categories, i.e. 1) First-order phase transition: $\frac{\partial A}{\partial F}\big|_{F\to 0}$ and $\frac{\partial A}{\partial T}\big|_{F\to 0}$ have discontinuities in a certain $T$ range; 2) Second-order phase transition: $\frac{\partial A}{\partial F}\big|_{F\to 0}$ and $\frac{\partial A}{\partial T}\big|_{F\to 0}$ are continuous with $T$, while $\frac{\partial^2 A}{\partial T \partial F}\big|_{F\to 0}$ and $\frac{\partial^2 A}{\partial T^2}\big|_{F\to 0}$ have discontinuities at a certain $T$ point; and 3) Diffuse phase transition: $\frac{\partial^3 A}{\partial T^2 \partial F}\big|_{F\to 0}$ and $\frac{\partial^3 A}{\partial T^3}\big|_{F\to 0}$ are continuous with $T$, while they are respectively equal to 0 at the transition-temperature ($T_d$) and diffuse-temperature ($T_s$), as well as the diffuse-region of the phase transition is $T_d$ to $T_s$ and the diffuse-degree is $\frac{T_s - T_d}{T_d}$, naturally giving the the relation of the phase transition to the specific heat peak. Deep analyses indicate that the classification of this paper is based on both the long-range and short-range correlation orders of the systems, while the Ehrenfest's or Landau's independently on the short-range or long-range order, respectively.

## Introduction

Relaxor-ferroelectrics[1-3] and relaxor-ferromagnets[4-6] have attracted a lot of attentions due to their unique physical properties and phase transition mechanisms[7,8]. Among them, there are the following contradictions in the classification of their phase transitions (referring to the thermal equilibrium phase transitions of thermodynamic limit systems). Due to the measured specific heat ($C$) being a diffuse peak[9,10] with temperature ($T$), and according to the conventional Ehrenfest classification of phase transitions[11], which takes the discontinuity or singularity of the derivative of the free energy ($A$) with $T$ ($\frac{\partial A}{\partial T}$ and $\frac{\partial^2 A}{\partial T^2}$ etc., where $C = T\frac{\partial^2 A}{\partial T^2}$) as the criterion ($A$ is always continuous with $T$), the above systems do not have any phase transition. However, relaxor-ferroelectrics and relaxor-ferromagnets have spontaneous polarization[3,12] and spontaneous magnetization[4,6,10] at low temperatures, indicating the existence of some kind of phase transitions, commonly known as diffuse phase transitions[5,8].

It should be noted that if Ehrenfest classification was extended as follows, specifically the criterion of diffuse phase transition is chosen as that $\frac{\partial^2 A}{\partial T^2}$ is continuous with $T$ and it is equal to 0 at a certain $T$ point, i.e. corresponding to a diffuse peak of $C$, then it can solve the problem of phase transition classification for relaxor-ferroelectrics and relaxor-ferromagnets, but it also produces the following contradictory result. For example, one-dimensional Ising model[13,14] exhibits a diffuse peak of $C$ vs $T$, and according to the above extended classification, there will be a diffuse phase transition at $0.83 J/k_B$, where $k_B$ is Boltzmann constant. However, its spontaneous magnetization is always 0 for $T > 0$, and it is generally believed that there is no phase transition at nonzero temperature. Therefore, it is not feasible to classify the diffuse phase transition solely based on the derivatives of $A$ with $T$.

In fact, the temperature corresponding to the minimum derivative of order parameter ($\eta$) to $T$ is generally used as the phase transition temperature in the characterization of diffusion phase transition[8,10,15,16]. Due to the fact that $\eta$ was first proposed by Landau[15,17], this method is referred to as Landau classification here. However, if it was believed that the Landau's is only applicable to diffuse phase transition, while the Ehrenfest's to first and second-order phase transitions, there exists the problem that the classifications are inconsistent; and if the Landau's was considered applicable to all types of phase transitions, i.e. abandoning the entropy ($S = -\frac{\partial A}{\partial T}\Big|_{F\to 0}$) and specific heat criteria of the Ehrenfest's, then a question that the relation of the specific heat peaks[9,10] above and near the transition temperature to the phase transitions is created. It should be noted that due to $\eta = -\left.\frac{\partial A}{\partial F}\right|_{F\to 0}$, Landau classification is based on $\left.\frac{\partial A}{\partial F}\right|_{F\to 0}$ and its derivatives to $T$ as the criteria.

In response to the above questions, as well as the benefits of scientific classification to phase transitions in discovering the relationships and laws between the macroscopic phenomena and microscopic mechanisms, this article reclassifies the thermal equilibrium phase transitions of thermodynamic limit systems, i.e. a new classification of phase transitions is proposed.

## Results

### Reclassification of phase transitions

Here, we simultaneously take $\left.\frac{\partial A}{\partial F}\right|_{F\to 0}$, $\left.\frac{\partial A}{\partial T}\right|_{F\to 0}$, and their derivatives to $T$ as the criteria to classify the thermal equilibrium phase transitions of thermodynamic limit systems into three categories, i.e. first-order, second-order, and diffuse phase transitions:

1. *First-order phase transition*: $\left.\frac{\partial A}{\partial F}\right|_{F\to 0}$ and $\left.\frac{\partial A}{\partial T}\right|_{F\to 0}$, i.e. $\eta$ and $S$, both have discontinuities in the temperature range of $T_1$ to $T_2$ ($T_1 < T_2$).

   $T_1$ and $T_2$ is called the lower transition temperature and the upper transition temperature, respectively. The states below $T_1$ and above $T_2$ are called ordered phase (e.g. ferroelectric or ferromagnetic phase) and disordered phase (e.g. paraelectric or paramagnetic phase), respectively. The temperature region from $T_1$ to $T_2$ is called the coexistence-region of two phases, and the temperature point where the two phases have equal $A$ is called the first-order phase transition temperature ($T_0$).

2. *Second-order phase transition*: $\left.\frac{\partial A}{\partial F}\right|_{F\to 0}$ and $\left.\frac{\partial A}{\partial T}\right|_{F\to 0}$ is continuous with temperature, while $\left.\frac{\partial^2 A}{\partial T \partial F}\right|_{F\to 0}$ and $\left.\frac{\partial^2 A}{\partial T^2}\right|_{F\to 0}$, i.e. $\frac{\partial \eta}{\partial T}$ and $C$, have discontinuities at a certain temperature point ($T_c$).

   The states below and above $T_c$ are called ordered and disordered phases, respectively, and $T_c$ is called the second-order phase transition temperature or critical temperature.

3. *Diffuse phase transition*: $\left.\frac{\partial^3 A}{\partial T^2 \partial F}\right|_{F\to 0}$ and $\left.\frac{\partial^3 A}{\partial T^3}\right|_{F\to 0}$, i.e. $\frac{\partial^2 \eta}{\partial T^2}$ and $\frac{\partial C}{\partial T}$, is continuous with temperature, but they are equal to zero at different temperature points $T_d$ and $T_s$ ($T_d < T_s$).

   $T_d$ and $T_s$ is referred to as the transition temperature and diffuse temperature, respectively. The states below $T_d$ and above $T_s$ are called ordered and disordered phases, respectively; the temperature region between $T_d$ and $T_s$ is the diffuse-region; and the diffuse-degree ($\delta$) is $\delta \equiv \frac{T_s - T_d}{T_d}$.

Table 1 New classification of thermal equilibrium phase transitions in thermodynamic limit systems

|  | First-order | Second-order | Diffuse |
|---|---|---|---|
| $\left.\frac{\partial A}{\partial F}\right\|_{F\to 0}$ i.e. $\eta$ | discontinuous in a temperature range of $T_1$ to $T_2$ | continuous with temperature |  |
| $\left.\frac{\partial A}{\partial T}\right\|_{F\to 0}$ i.e. $S$ |  |  |  |
| $\left.\frac{\partial^2 A}{\partial T \partial F}\right\|_{F\to 0}$ i.e. $\frac{\partial \eta}{\partial T}$ |  | discontinuous at a temperature point ($T_c$) |  |
| $\left.\frac{\partial^2 A}{\partial T^2}\right\|_{F\to 0}$ i.e. $C$ |  |  |  |
| $\left.\frac{\partial^3 A}{\partial T^2 \partial F}\right\|_{F\to 0}$ i.e. $\frac{\partial^2 \eta}{\partial T^2}$ |  |  | continuous with temperature, and equal to zero at a temperature point ($T_d$) |
| $\left.\frac{\partial^3 A}{\partial T^3}\right\|_{F\to 0}$ i.e. $\frac{\partial C}{\partial T}$ |  |  | continuous with temperature, and equal to zero at a temperature point ($T_s$) |

The simple inductions of the new classification is shown in Table 1, and some additional explanations to the new classification of this article are:

1) The classification only applies to thermodynamic limit systems, where the molecule number is approximately the Avogadro constant, rather than microscopic systems that contain a very small number of molecules.

2) The classification only focuses on thermal equilibrium phase transitions, and does not include non-thermal equilibrium glass transitions[17].

3) Due to the fact that the third and higher order phase transitions according to Ehrenfest classification have not been confirmed experimentally, they have not been classified here.

4) The measurement methods for spontaneous polarization ($P_s$) include the electric hysteresis-loop and heating-pyroelectric after strong-electric-field-cooling[12]. The methods for spontaneous magnetization ($M_s$) have the magnetic hysteresis-loop, Arrott-plot[19], and weak-field-heating after strong-field-cooling[16].

5) In experiments, if the adjacent derivative of the measured values differs by more than one order of magnitude, it is generally referred to as a discontinuity.

6) Measured specific heat usually includes background components unrelated to phase transitions, such as the contribution of phonons. The $C$ in this article refers to the specific heat deducted this part.

## Examples of phase transition reclassification

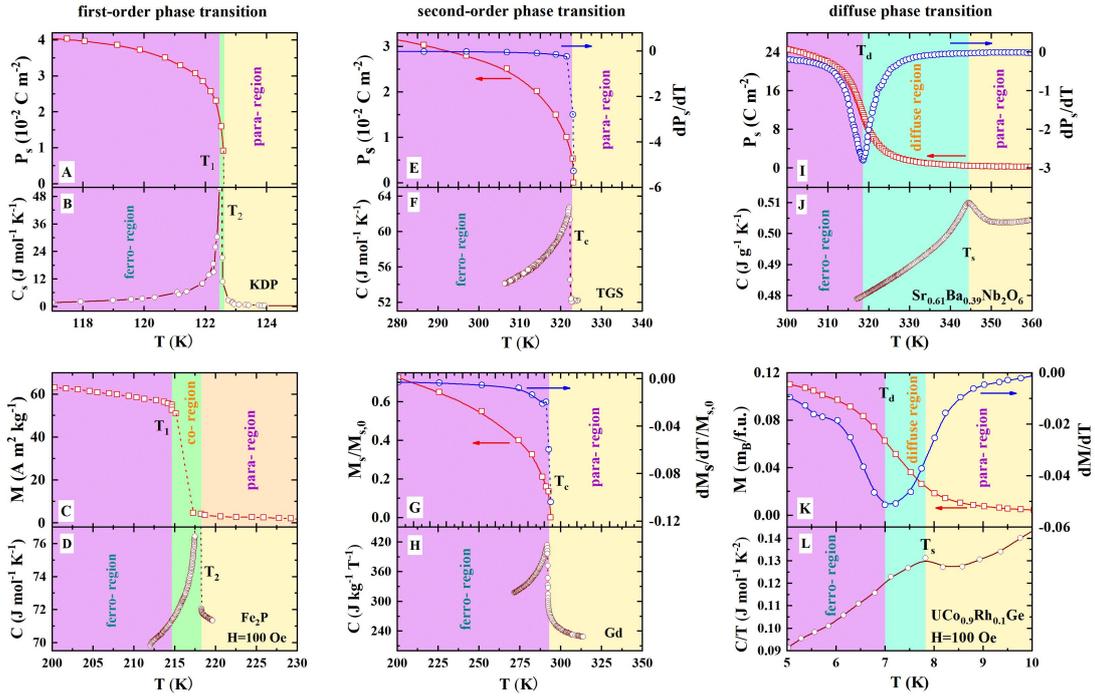

Figure 1 *First-order phase transition*: (A) Experimental results (red square) of spontaneous polarization ($P_s$) of ferroelectric potassium dihydrogen phosphate (KDP) single crystal vs temperature ($T$)[20]. The light-purple and light-yellow regions represent ferroelectric phase below the lower transition temperature ($T_1$) and paraelectric phase above the upper transition temperature ($T_2$), respectively. The light-green region represents the ferroelectric and paraelectric coexistence-region from $T_1$ to $T_2$. The red dashed line indicates the discontinuity of $P_s$ in the coexistence-region. (B) Experimental results (wine-red circle) of the subtracted background specific heat ($C_s$) of KDP single crystal vs $T$ (without external field)[21]. The wine-red dashed line indicates the divergence in the coexistence-region. (C) Experimental results (red square) of the magnetization ($M$) of ferromagnetic Fe$_2$P single crystal with $T$, measured magnetic field $H = 100\ Oe$[22]. The light-purple and light-yellow regions represent the ferromagnetic phase lower than $T_1$ and paramagnetic phase higher than $T_2$. The light-green region indicates the ferromagnetic and paramagnetic coexistence-region from $T_1$ to $T_2$. The red dashed line indicates the discontinuity of $M$ in the coexistence-region. (D) Experimental results (wine-red circle) of the specific heat ($C$) of Fe$_2$P single crystal vs $T$ (without external field)[22]. The wine-red dashed line indicates the divergence of $C$ in the coexistence-region.

*Second-order phase transition*: (E) Experimental results (red square) of $P_s$ and $dP_s/dT$ (blue circle) of ferroelectric triglycine sulfate (TGS) single Crystal vs $T$ [23]. The blue dashed line represents the discontinuity of $dP_s/dT$ at the phase transition temperature ($T_c$). (F) Experimental results of $C$ (wine-red circle) of TGS single crystal vs $T$ (without external field)[24]. The wine-red dashed line indicates the discontinuity of $C$ at $T_c$. (G) Experimental results of the reduced spontaneous magnetization ($M_s/M_{s,0}$, red square) and $dM_s/dT/M_{s,0}$ (blue circle) of ferromagnetic gadolinium (Gd) single crystal vs $T$[25]. The blue dashed line represents the discontinuity of $dM_s/dT$ at $T_c$. (H) Experimental results of $C$ (wine-red circle) of Gd single crystal vs $T$ (without external field)[26]. The wine-red dashed line marks the discontinuity of $C$ at $T_c$.

*Diffuse phase transition*: (I) Experimental results of $P_s$ (red square) and $dP_s/dT$ (blue circle) of relaxor-ferroelectric Sr$_{0.61}$Ba$_{0.39}$Nb$_2$O$_6$ single crystal vs $T$[12]. The transition temperature ($T_d$) corresponds to the minimum value of $dP_s/dT$. (J) Experimental results of $C$ (wine-red circle) of Sr$_{0.61}$Ba$_{0.39}$Nb$_2$O$_6$ single crystal vs $T$ (without external field)[9]. The diffuse temperature ($T_s$) is the peak temperature of $C$. The cyan region from $T_d$ to $T_s$ is the diffuse-region of phase transition. (K) Experimental results of $M$ (red square) and $dM/dT$ (blue circle) of relaxor-ferromagnetic UCo$_{0.9}$Rh$_{0.1}$Ge single crystal vs $T$, measured $H = 100\ Oe$[10]. $T_d$ corresponds to the minimum value of $dM/dT$. (L) Experimental Results of $C$ (wine-red circle) of UCo$_{0.9}$Rh$_{0.1}$Ge single Crystal vs $T$[10]. $T_s$ is the peak temperature of $C$.

Here, we takes the phase transitions of ferroelectric and ferromagnetic systems as examples, including: 1) Macroscopic single crystals with homogeneous distribution of components[20-23]; 2) Macroscopic single crystals with heterogeneous distribution of components[9,10,12]; and 3) A special case that has long troubled phase transition classification, i.e. one-dimensional Ising model[13,14].

*First-order phase transition*: Fig. 1A and B respectively show the experimental results of spontaneous polarization ($P_s$, i.e. $\eta$)[20], and specific heat ($C_s$, without external field)[21], subtracting the background of potassium dihydrogen phosphate (KDP) single crystal, a ferroelectric of homogeneous distribution of components, vs $T$. $P_s$ exhibits a discontinuity near 112.62 K, while $C_s$ diverges between 112.46 and 112.58 K, resulting in $T_1 \approx 112.46$ K and $T_2 \approx 112.62$ K. Fig. 1C and D respectively give the experimental results of the magnetization ($M$, i.e. $\eta$, measured magnetic field $H = 100$ Oe) and $C$ ($H = 0$) of Fe$_2$P single crystal[22], a ferromagnet of homogeneous distribution of components, vs $T$. $M$ has a discontinuity between 214.7 and 217.3 K, and $C$ diverges between 214.7 and 217.9 K, indicating $T_1 \approx 214.7$ K and $T_2 \approx 217.9$ K.

*Second-order phase transition*: Fig. 1E and F show the experimental results of $P_s$[23], $\frac{dP_s}{dT}$ and $C$ (without external field)[24] of triglycine sulfate (TGS) single crystal, a ferroelectric of homogeneous distribution of components, vs $T$. $P_s$ is continuous with $T$, while $\frac{dP_s}{dT}$ and $C$ have discontinuities at $T_c = 323.1$ K. Fig 1G and H give the experimental results of the reduced spontaneous magnetization ($\frac{M_s}{M_{s,0}}$)[25], $\frac{dM_s}{dT}/M_{s,0}$ and $C$[26] of ferromagnetic gadolinium (Gd) single crystals vs $T$, where $M_{s,0}$ is the value of $M_s$ for $T \to 0$. $M_s$ is continuous with $T$, but $\frac{dM_s}{dT}$ and $C$ is discontinuous at $T_c = 293.5$ K.

The above experimental results indicate that the classification proposed in this paper is consistent with the Ehrenfest's and Landau's for the first and second-order phase transitions.

*Diffuse phase transition*: Fig. 1I and J show the experimental results of $P_s$[12], $\frac{dP_s}{dT}$ and $C$[9] of Sr$_{0.61}$Ba$_{0.39}$Nb$_2$O$_6$ single crystal, a relaxor-ferroelectric of heterogeneous distribution of components, vs $T$. $\frac{dP_s}{dT}$ and $C$ are continuous with $T$, while they have a minimum ($\frac{d^2P_s}{dT^2} = 0$) and a maximum ($\frac{dC}{dT} = 0$) at 318.6 K and 323.2 K, respectively. According to the classification of this article, Sr$_{0.61}$Ba$_{0.39}$Nb$_2$O$_6$ undergoes a diffuse phase transition at $T_d = 318.6$ K, $T_s = 323.2$ K, and $\delta = 1.44\%$. Fig. 1K and L gives the experimental results of $M$ ($H = 100$ Oe), $\frac{dM}{dT}$ and $C$ of UCo$_{0.9}$Rh$_{0.1}$Ge single crystal[10], a relaxor-ferromagnet of heterogeneous distribution of components, vs $T$. $\frac{dM}{dT}$ and $C$ are continuous with $T$, and show a minimum ($\frac{d^2M}{dT^2} = 0$) and a maximum ($\frac{dC}{dT} = 0$) at $T_d = 7.00$ K and $T_s = 7.78$ K, respectively, and the corresponding $\delta = 11.1\%$.

Especially, based on the classification of this article, a special case that has long troubled phase transition classification, i.e. one-dimensional Ising model, which gives different results according to the Ehrenfest's and Landau's, is reasonably classified. Specifically, the model does not have any phase transition[13,14] by the Ehrenfest's, while based on the Landau's, this model has a phase transition of $T \to 0$ K, commonly known as a zero-temperature phase transition[27]. According to our classification, the model exists a diffusion phase transition of $T_d \to 0$ K, $T_s = 0.83 J/k_B$, and $\delta \to \infty$, an extreme case of diffuse phase transitions.

Therefore, the classification of this article not only overcomes the difficulty of the Ehrenfest's (Sr$_{0.61}$Ba$_{0.39}$Nb$_2$O$_6$, UCo$_{0.9}$Rh$_{0.1}$Ge and one-dimensional Ising mode have no phase transition), but also the question of the Landau's (the connection between the phase transition and specific heat peak above and near $T_d$ in Sr$_{0.61}$Ba$_{0.39}$Nb$_2$O$_6$, UCo$_{0.9}$Rh$_{0.1}$Ge and one-dimensional Ising mode), as well as naturally defines the diffuse-region and diffuse-degree of the transition.

## Discussion

For real macroscopic systems of phase transitions are electrically neutral, the total interaction between dipoles due to electrostatic shielding is a short-range interaction, and the spatial scale of the interaction is about the order of crystal cell parameters. Therefore, the nearest-neighbor interaction model can describe the systems approximately. For clarity and without losing generality, the random-site Heisenberg model (RS-HM) of nearest-neighbor interactions[17] is used as an example to illustrate the classification of phase transitions in this paper (for the random bond Heisenberg model gives the spin glass transition[17], it is not discussed here). The Hamiltonian of this model is,

$$\mathcal{H} = -\frac{J}{2}\sum_{i \neq j}^{\{nn\}} \vec{\mu}_i \cdot \vec{\mu}_j r_i^\phi r_j^\phi \qquad (1)$$

where $\vec{\mu}_i$ is the moment of the dipole on the $i$-th lattice point in the crystal lattice of the system; $J$ the interaction constant between the nearest-neighbor dipoles; $\phi$ the concentration of the dipole ($0 < \phi \leq 1$); $r_i^\phi$ the distribution of dipoles on the lattice that is a random function of 1 or 0 determined by $\phi$, i.e. randomly generating a random number $r$ between 0 and 1, $r_i^\phi = 1$ if $r \leq \phi$, indicating there is a dipole on the lattice point, otherwise $r_i^\phi = 0$, which means no dipole or dipole vacancy on the lattice point; $\{nn\}$ represents the nearest-neighbor sum of $i$-th and $j$-th dipoles.

It should be noted that: 1) For strongly anisotropic systems, RS-HM is simplified as the random-site Ising model[8,17]; 2) When $\phi = 1$, RS-HM is the traditional Heisenberg model, which describes the second-order ferroelectric and ferromagnetic phase transitions in systems with homogeneous distribution of components; 3) When $\phi$ is neither very big nor very small, RS-HM is one of the most feasible models for describing the diffuse ferroelectric and ferromagnetic phase transitions in systems with heterogeneous distribution of components (the systems of smaller $\phi$ are generally believed to undergo glass transition)[8,17]; 4) The description of first-order ferroelectric and ferromagnetic phase transitions requires the modified term added to the Heisenberg or Ising models[28], but these modifications do not affect the following discussion[10].

Based on Boltzmann-Gibbs statistics[8,17] and according to Eq.1, the average internal energy ($U$) per dipole in the systems at thermal equilibrium can be obtained as,

$$U = -\lim_{N \to \infty} \frac{J}{2\phi N} \sum_{i=1}^{N} \zeta_i^{nn} \tag{1}$$

where $N$ is the number of crystal lattice points, $\zeta_i^{nn} \equiv \langle \vec{\mu}_i \cdot \vec{\mu}_j r_i^\phi r_j^\phi \rangle_{nn}$ that is the correlation function between nearest-neighbor dipoles, commonly referred to as the short-range correlation order, short-range order for short, of the systems[8,17] ($\langle \cdots \rangle$ is the thermodynamic statistical average of the orientation configurations of the dipoles).

From the relationships between $U$, $S$ and $C$ of the thermal equilibrium systems, it is easy to obtain that $S$ and $C$ are also the functions of $\zeta_i^{nn}$, i.e. related to the short-range order of the systems.

On the other hand, $\eta$ of thermal equilibrium systems can also be expressed as the following form[29],

$$\eta = \lim_{N \to \infty} \frac{1}{N\phi} \sqrt{\sum_{i \neq j = 1}^{N} \zeta_{i,j}} \tag{2}$$

where $\zeta_{i,j} \equiv \langle \vec{\mu}_i \cdot \vec{\mu}_j r_i^\phi r_j^\phi \rangle$ that is the correlation function between the $i$-th and $j$-th dipoles in the systems.

It can be seen that $\eta$ reflects $\zeta_{i,j}$ for $|i - j| \to \infty$, i.e. the long-range correlation function, commonly referred to as the long-range correlation order, long-range order for short, of the systems at mesoscopic (approximately $10^3$ cell constants) and even macroscopic scales[17,29].

Therefore, the phase transition classification in this article accords to the characteristics of both the short-range and long-range orders in systems vs temperature, i.e. double-order classification, while the Ehrenfest's or Landau's the short-range or long-range order independently, i.e. single-order classification.

For the three-dimensional uniform macroscopic systems (i.e. $\phi = 1$ and $r_i^\phi = 1$), such as KDP, Fe2P, TGS, Gd, etc., the experimental results (Fig. 1A-H) show that the formation temperature of the short-range order is the same as that of the long-range order.

For systems with heterogeneous distribution of components (i.e. $\phi < 1$), such as $Sr_{0.61}Ba_{0.39}Nb_2O_6$ and $UCo_{0.9}Rh_{0.1}Ge$, due to the fractal cluster characteristics of dipoles distributed on the crystal lattices[8,30,31], on the one hand, short-range order first form inside the clusters, while long-range order reflecting the correlation between clusters at lower temperatures[8,18] during the cooling process, i.e. $T_d < T_s$; On the other hand, it leads to the spatial distribution of short and long-range orders[30,31], which means that $C$ and $\eta$ are continuous and differentiable with temperature.

Because the thermal equilibrium phase transition of macroscopic systems is essentially the formation or disappearance process of the long-range order, this article defines $T_d$ as the transition temperature, which is consistent with the general definition in the literature[8,10,16]. As for $T_s$, it is generally referred to as the crossover temperature between the paramagnetic and correlated-paramagnetic in magnetic materials[10], i.e. the state above $T_s$ is paramagnetic, which is consistent with the definition that it is the diffusion temperature of phase transition in this article.

According to the double-order classification in this article (as shown in Fig. 1I-L): 1) The disordered (paraelectric or paramagnetic) phase above $T_s$ is a state that the short-range order is small while the long-range order tends to zero; 2) The state of the diffuse region between $T_d$ and $T_s$ is that the short-range order is large while the long-range order is small; and 3) The ordered (ferroelectric or ferromagnetic) phase below $T_d$ is a state that both the short-range and long-range orders are large. It could be imagined that the classification results are undoubtedly enlightening for exploring the microscopic mechanisms of diffuse phase transitions.

## References

1. Chen, L. et al. Giant energy-storage density with ultrahigh efficiency in lead-free relaxors via high-entropy design. *Nat. Comm.* **13**, 3089 (2022).
2. Petzelt, J. et al. Broadband Dielectric, Terahertz, and Infrared Spectroscopy of BaTiO3-BaZrO3 Solid Solution: From Proper Ferroelectric over Diffuse and Relaxor Ferroelectrics and Dipolar Glass to Normal Dielectric. *Phys. Stat. Sol. B*. **258**, 2100259 (2021).
3. Cross, L. E. Relaxor ferroelectrics. *Ferroelectrics*. **76**, 241-267 (1987).
4. Chin, C. et al. Magnetic properties of La3Ni2SbxTayNb1-x-yO9; from relaxor to spin glass. *J. Sol. Stat. Chem.* **273**, 175-185 (2019).
5. Kimura, T., Tomioka, Y., Kumai, R., Okimoto, Y. & Tokura, Y. Diffuse phase transition and phase separation in Cr-doped Nd1/2Ca1/2MnO3: A relaxor ferromagnet. *Phys. Rev. Lett.* **83**, 3940-3943 (1999).
6. Battle, P. D., Evers, S. I., Hunter, E. C. & Westwood, M. La3Ni2SbO9: a Relaxor Ferromagnet. *Inorg. Chem.* **52**, 6648-6653 (2013).
7. Pirc, R., Blinc, R. & Scott, J. F. Mesoscopic model of a system possessing both relaxor ferroelectric and relaxor ferromagnetic properties. *Phys. Rev. B*. **79**, 21411421 (2009).
8. Zhang, L. L. & Huang, Y. N. Theory of relaxor-ferroelectricity. *Sci. Rep.* **10**, 5060 (2020).
9. Kleemann, W., Dec, J., Shvartsman, V. V., Kutnjak, Z. & Braun, T. Two-dimensional Ising model criticality in a three-dimensional uniaxial relaxor ferroelectric with frozen polar nanoregions. *Phys.*

*Rev. Lett.* **97**, 65702 (2006).

10. Pospisil, J. et al. Intriguing behavior of UCo1-xRhxGe ferromagnets in magnetic field along the b axis. *Phys. Rev. B*. **102**, (2020).
11. Ehrenfest, P. Phase conversions in a general and enhanced sense, classified according to the specific singularities of the thermodynamic potential. *Proc. Konin. Aka. Weten. Amst.* **36**, 153-157 (1933).
12. Granzow, T., Woike, T., Wohlecke, M., Imlau, M. & Kleemann, W. Change from 3D-Ising to random field-Ising-model criticality in a uniaxial relaxor ferroelectric. *Phys. Rev. Lett.* **92**, 65701 (2004).
13. Ising, E. Report on the theory of ferromagnetism. *Z. Phys.* **31**, 253-258 (1925).
14. McCoy, B. M. & Wu, T. T. *The two-dimensional Ising model*. (Dover Publications, Inc., Mineola, New York, 2014).
15. Landau, L. The theory of phase transitions. *Nature*. **138**, 840-841 (1936).
16. Huy, N. T. & de Visser, A. Ferromagnetic order in U(Rh,Co)Ge. *Sol. Stat. Comm.* **149**, 703-706 (2009).
17. Binder, K. & Young, A. P. Spin-glasses: Experimental facts, theoretical concepts, and open questions. *Rev. Mod. Phys.* **58**, 801-976 (1986).
18. Arrott, A. Criterion for ferromagnetism from observations of magnetic isotherms. *Phys. Rev.* **108**, 1394-1396 (1957).
19. Benepe, J. W. & Reese, W. Electronic studies of KH2PO4. *Phys. Rev. B*. **3**, 3032-3039 (1971).
20. Strukov, B. A., Amin, M. & Kopchik, V. A. Comparative investigation of specific heat of KH2PO4 (KDP) and KD2PO4 (DKDP) single crystals. *Phys. Stat. Sol.* **27**, 741-749 (1968).
21. Hudl, M. et al. Thermodynamics around the first-order ferromagnetic phase transition of $Fe_2P$ single crystals. *Phys. Rev. B*. **90**, 144432 (2014).
22. Shibuya, I. & Mitsui, T. Ferroelectric phase transition in (glycine)3H2SO4 and critical x-ray scattering. *J. Phys. Soc. Jpn.* **16**, 479-489 (1961).
23. Gallardo, M. C., Martin-Olalla, J. M., Romero, F. J., Del Cerro, J. & Fugiel, B. Memory effect in triglycine sulfate induced by a transverse electric field: specific heat measurement. *J. Phys.-Cond. Mat.* **21**, 259022 (2009).
24. Nigh, H. E., Legvold, S. & Spedding, F. H. Magnetization and electrical resistivity of gadolinium single crystals. *Phys. Rev.* **132**, 1092-1097 (1963).
25. Glorieux, C., Thoen, J., Bednarz, G., White, M. A. & Geldart, D. J. Photoacoustic investigation of the temperature and magnetic-field dependence of the specific-heat capacity and thermal conductivity near the Curie point of gadolinium. *Phys. Rev. B*. **52**, 12770-12778 (1995).
26. Choi, D. et al. *Colloquium*: Atomic spin chains on surfaces. *Rev. Mod. Phys.* **91**, (2019).
27. Yin, H., Zhou, H. & Huang, Y. A New Model of Ferroelectric Phase Transition with Neglectable Tunneling Effect. *Chin. Phys. Lett.* **36**, 70501 (2019).
28. Schultz, T. D., Mattis, D. C. & Lieb, E. H. 2-dimensional Ising model soluble problem of many fermions. *Rev. Mod. Phys.* **36**, 856-871 (1964).
29. Shvartsman, V. V., Dkhil, B. & Kholkin, A. L. Mesoscale domains and nature of the relaxor state by piezoresponse force microscopy. *Ann. Rev. Mat. Res.* **43**, 423-449 (2013).
30. Shvartsman, V. V. & Lupascu, D. C. Lead-free relaxor ferroelectrics. *J. Am. Cera. Soc.* **95**, 1-26 (2012).
31. Ferdinand, A. E. & Fisher, M. E. Bounded and inhomogeneous Ising models: I. specific-heat

anomaly of a finite lattice. *Phys. Rev.* **185**, 832-846 (1969).